\documentclass[lettersize,journal]{IEEEtran}
\ifCLASSINFOpdf
\else
   \usepackage{graphicx}
   \graphicspath{{../eps/}}
   \DeclareGraphicsExtensions{.eps}
\fi

\usepackage{amsmath}
\usepackage{amsfonts,amssymb}
\usepackage{amssymb}
\usepackage{wrapfig}
\usepackage{psfrag}
\usepackage{epstopdf}
\usepackage{cite}
\usepackage{graphicx}
\usepackage{float} 
\usepackage{subfigure}
\usepackage{threeparttable}
\usepackage{cases}
\usepackage{subeqnarray}
\usepackage{color}
\usepackage{underscore}
\usepackage{verbatim}
\usepackage{bm}
\usepackage{stfloats}
\usepackage{ragged2e} 
\usepackage{diagbox}
\usepackage{multirow}
\usepackage{longtable}
\usepackage{url}
\usepackage[normalem]{ulem}
\usepackage{tabularray}
\newtheorem{theorem}{Theorem}

\newtheorem{assumption}{Assumption}

\usepackage{algorithm}
\usepackage{algorithmic}

\begin{document}
\title{Can Channels be Fully Inferred Between

  Two Antenna Panels?}
%
%
%

\author{Yuelong~Qiu, Di~Wu, 
        and
        Yong~Zeng,~\IEEEmembership{Senior Member,~IEEE}
\thanks{This work was supported by the National Key R\&D Program of China with grant number 2019YFB1803400, in part by the National Natural Science Foundation of China with grant number 62071114.}
\thanks{The authors are with the National Mobile Communications Research Laboratory, Southeast University, Nanjing 210096, China.
 
Yong Zeng is also with the Purple Mountain Laboratories, Nanjing 211111, China (e-mail: {yl_qiu, studywudi, yong_zeng}@seu.edu.cn). }
}

\maketitle

\begin{abstract}
 This letter considers a two-panel massive multiple-input multiple-output (MIMO) communication system, where the base station (BS) is equipped with two antenna panels that may use different frequency bands for communication. By exploiting the geometric relationships between antenna panels, eff{}icient channel inference methods across antenna panels are proposed to reduce the overhead of real-time channel estimation. Four scenarios are considered, namely far-f{}ield free-space, near-f{}ield free-space, multi-path sharing far-f{}ield scatterers, and multi-path sharing near-f{}ield scatterers. For both far-f{}ield and near-f{}ield free-space scenarios, we show that the channel of one panel can be fully inferred from that of the other panel, as long as the multi-path components (MPCs) composing the channel can be resolved. On the other hand, for the multi-path scenarios sharing far-f{}ield or near-f{}ield scatterers, only the angles or range of angles of the MPCs can be inferred, respectively. Simulation results based on commercial 3D ray-tracing software are presented to validate our developed channel inference techniques.
\end{abstract}


\IEEEpeerreviewmaketitle
\vspace{-4mm}
\section{Introduction}
Massive multiple-input multiple-output (MIMO) is one of the key technologies for enabling millimeter wave (mmWave) communications \cite{heath2016overview}. To reduce the hardware cost and power consumption of massive MIMO systems, the 3rd Generation Partnership Project (3GPP) has proposed the use of antenna in package (AIP) technology. Specif{}ically, AIP juxtaposes multiple antenna panels to form multi-panel massive MIMO \cite{huawei2016antenna}. As a partially-connected hybrid architecture, each antenna panel in the multi-panel massive MIMO system is linked to a modest number of radio frequency (RF) chains, resulting in signif{}icant reductions in hardware costs and energy consumption \cite{wang2020orthogonal}. This has made the system increasingly popular in the industry, especially in high-frequency bands \cite{CATT1702071}. However, the large antenna array composed of multiple antenna panels are typically non-uniform in structure, meaning that the distance between adjacent panels is much greater than the distance between adjacent antenna elements \cite{huawei2016antenna}. The cost of channel state information (CSI) acquisition is increased in multi-panel massive MIMO systems due to two factors. F{}irstly, the presence of multiple antenna panels requires the acquisition of more CSI as compared to one single panel. Secondly, due to the large inter-panel separations, different antenna panels may have different channel parameters, such as number of multi-paths, angles of arrivals/departures (AoAs/AoDs). A potential solution to reduce CSI acquisition overhead is channel extrapolation, which tries to infer the CSI of one antenna panel based on that of another one.

There have been extensive studies on channel extrapolation, which can be classif{}ied into time-, frequency-, and spatial-domain extrapolation. For time-domain channel extrapolation, current research mainly focuses on time division duplexing (TDD) systems \cite{zhang2022deep}. For frequency-domain channel extrapolation, some studies tried to achieve extrapolation within Sub-6 GHz or from Sub-6 GHz to mmWave \mbox{\cite{alrabeiah2019deep},\cite{ali2019spatial}}. For spatial-domain channel extrapolation, most research mainly focuses on channel extrapolation between different antennas on the same panel \cite{lin2021deep}, while channel extrapolation between different base stations (BSs) and different panels is rare \cite{chen2017remote}, \cite{wang2018channel}. Furthermore, artif{}icial intelligence (AI)-based time-, frequency- and spatial-domain channel extrapolation is discussed in \cite{zhang2023ai}. Additionally, a novel concept termed channel knowledge map (CKM) was recently proposed \cite{zeng2021toward}, \cite{zeng2023tutorial}, which can enhance environment-awareness and facilitate CSI acquisition by utilizing location and environment information.

In this letter, we consider a two-panel massive MIMO communication system, where the BS is equipped with two antenna panels for communication. The main objective is to infer the channel of one panel using that of the other panel. Unlike the multi-panel scenario considered by 3GPP, where the antenna panels mainly operate in the same frequency band \cite{huawei2016antenna}, \cite{CATT1702071}, we consider the more general scenario that the panels may use different frequency bands. Moreover, since the separation between panels is larger than that between antenna elements within a panel, channel inference across antenna panels is more challenging than channel extrapolation between antenna elements on the same panel. Fortunately, by exploiting the geometric structure between antenna panels at the BS, we show that the channel of one panel can be in principle fully inferred from that of the other panel for both far-f{}ield and near-f{}ield free-space scenarios, as long as the multi-path components (MPCs) composing the channel can be resolved. On the other hand, for the multi-path scenarios sharing far-f{}ield or near-f{}ield scatterers, only the angles or range of angles of the MPCs can be inferred, respectively. Numerical results based on commerical ray tracing software are provided to validate our analysis. 
\vspace{-2mm}
\section{System Model}
\begin{figure}[htbp]
\centering
\includegraphics[scale=0.45]{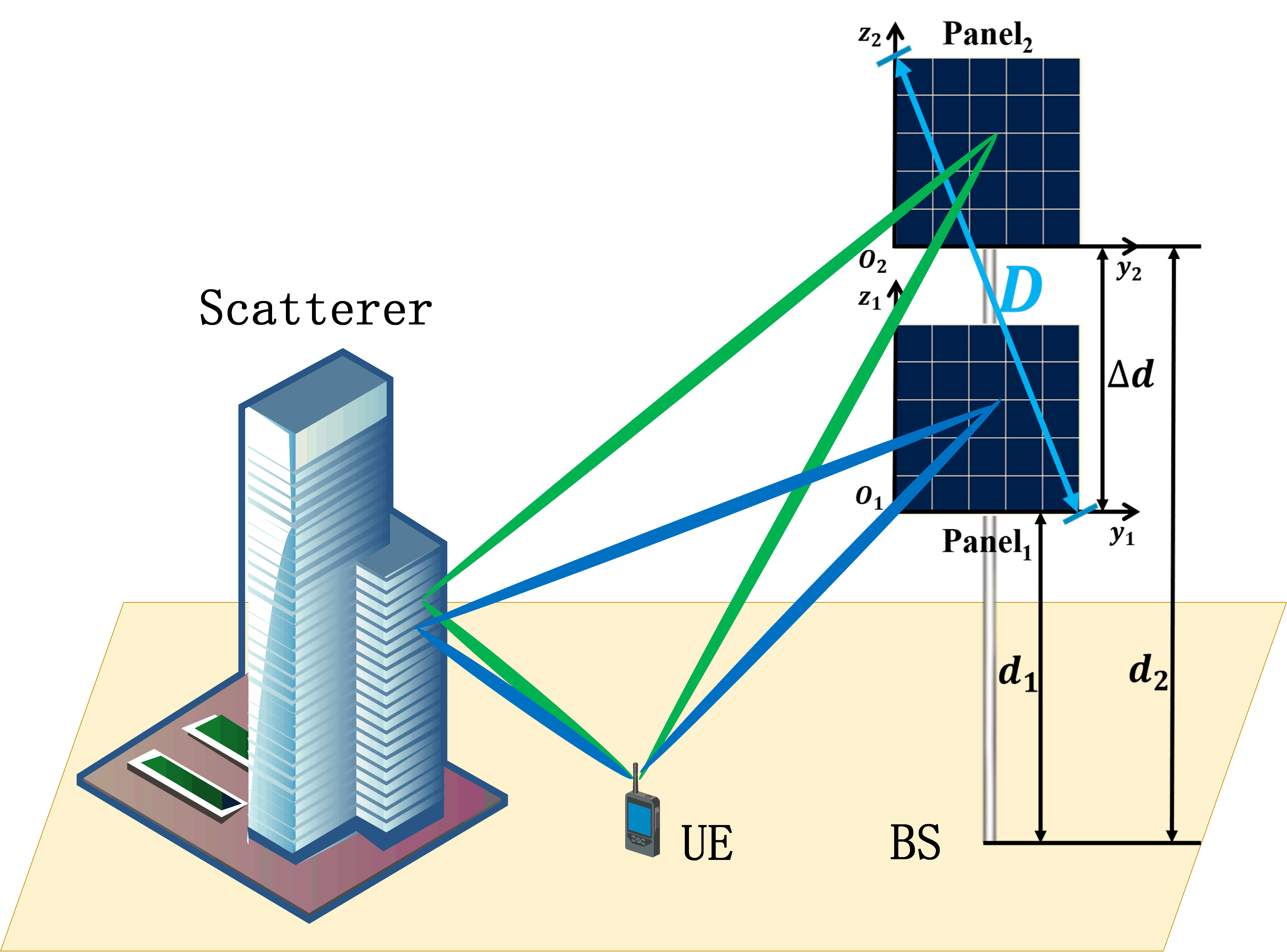}
\vspace{-2mm}
\caption{Wireless communication where the BS is equipped with two antenna panels.}
\label{fig1}
\end{figure}

\vspace{-2mm}
As shown in Fig.~\ref{fig1}, we consider a downlink wireless \mbox{communication} system where the BS is equipped with two antenna panels, denoted as ${\rm Panel}_{1}$ and ${\rm Panel}_{2}$, and their operating frequencies are denoted as ${{f}_{1}}$ and ${{f}_{2}}$, respectively. ${\rm Panel}_{1}$ and ${\rm Panel}_{2}$ are placed in the same vertical plane, with their bottom edges having heights $d_1$ and $d_2$, respectively.  Both ${\rm Panel}_{1}$ and ${\rm Panel}_{2}$ are uniform planar arrays (UPAs) with adjacent antenna elements separated by half wavelength. The bottom-left antenna elements of ${\rm Panel}_{1}$ and ${\rm Panel}_{2}$ are chosen as their reference elements, respectively. The total number of antenna elements in ${\rm Panel}_{1}$ is \mbox{${N}_{1}={{N}_{z_1}}{{N}_{y_1}}$}, where ${{N}_{z_1}}$ and ${{N}_{y_1}}$ are the number of elements in the z-axis and y-axis, respectively. Similarly, the number of antenna elements in ${\rm Panel}_{2}$ is \mbox{${N}_{2}={{N}_{z_2}}{{N}_{y_2}}$}. Furthermore, the distance between the reference antenna elements of ${\rm Panel}_{1}$ and ${\rm Panel}_{2}$ is denoted as $\Delta d$, and the overall maximum dimension formed by ${\rm Panel}_{1}$ and ${\rm Panel}_{2}$ is denoted by $D$, as illustrated in Fig.~\ref{fig1}.
\vspace{-6mm}

\vspace{-3mm}
We assume that the user equipment (UE) and scatterers are at the far-f{}ield of each panel, but may be located in the near-f{}ield of the overall aperture $D$ formed by the two panels. Therefore, the channel vectors between ${\rm Panel}_{k}$ and a single-antenna UE can be expressed as
\setlength\abovedisplayskip{3pt}
\setlength\belowdisplayskip{3pt}
\begin{equation}
{{\mathbf{h}}_{k}}=\sqrt{{{N}_{k}}}\sum\limits_{l=1}^{{{L}_{k}}}{\alpha _{l}\left( {{\lambda}_{k}} \right)}{{\bf{a}}_{k}}\left( \theta _{l,k},\phi _{l,k} \right), k=1, 2,\label{eq1}
\end{equation}
where ${L}_{k}$ is the total number of paths, ${{\alpha }_{l}}\left( {\lambda}_{k}  \right)$ is the complex gain of path $l$ whose value depends on the carrier wavelength ${\lambda}_{k}=c/{f}_{k}$, with $c$ being the speed of light and $f_k$ being the frequency, ${\bf{a}}_{k}(\cdot)$ is the steering vector of the antenna array at ${\rm Panel}_{k}$, and $\theta_{l,k}\in\left [-\frac{\pi}{2}, \frac{\pi}{2}  \right ] $ and $\phi  _{l,k}\in\left [ 0,2\pi \right ) $ are the elevation and azimuth angles of path $l$ of ${\rm Panel}_{k}$, respectively.

With adjacent elements separated by half-wavelength for both antenna panels, the array steering vector for angle ($\theta$, $\phi$) can be expressed as
\begin{equation}
{{\bf{a}}_{k}}\left( \theta,\phi \right)={{\bf{a}}_{y_k}}\left( \theta,\phi \right)\otimes {{\bf{a}}_{z_k}}\left( \theta \right),\label{eq2}
\end{equation}
where
\begin{equation}
\begin{split}
{{\bf{a}}_{y_k}}\left( \theta ,\phi \right)= \frac{1}{\sqrt{{{N}_{y_k}}}}&[ 1,{{e}^{j\pi \cos ( {{\theta }} )\sin ( {{\phi }} )}},\cdots ,\\
&{{e}^{j\pi \left( {{N}_{y_k}}-1 \right)\cos ( {{\theta }} )\sin ( {{\phi }} )}}]^T,\label{eq3}
\end{split}
\end{equation}
\begin{equation}
{{\bf{a}}_{z_k}}\left( \theta \right)=\frac{1}{\sqrt{{{N}_{z_k}}}}[ 1,{{e}^{j\pi \sin ( {{\theta}} )}},\cdots ,{{e}^{j\pi \left( {{N}_{z_k}}-1 \right)\sin ( {{\theta }} )}} ]^T.\label{eq4}
\end{equation}

\vspace{-1mm}
For the same UE, each antenna panel needs to obtain its own CSI. To reduce the CSI acquisition overhead for \mbox{two-panel} systems, we exploit the geometric relationship between panels to achieve channel inference from ${\mathbf{h}}_{1}$ to ${\mathbf{h}}_{2}$. As such, only CSI of ${\mathbf{h}}_{1}$ needs to be estimated in real-time. To that end, we make the following assumption:

\begin{assumption}
The multi-path parameters ${{\alpha }_{l}}\left( {\lambda}_{1}  \right)$, $\theta _{l,1}$, and $\phi _{l,1}, l=1,...,L_{1}$, can be obtained based on the channel vector ${{\mathbf{h}}_{1}}$.\label{Assumption 1}
\end{assumption}

Assumption 1 is valid if ${\rm Panel}_{1}$ has a suff{}iciently large antenna number ${N}_{1}$ and/or the channel is sparse, so that $L_{1}\ll N_{1}$. In this case, one may use algorithms like multiple signal classif{}ication (MUSIC) \cite{schmidt1986multiple}, space-alternating generalized expectation-maximization (SAGE) \cite{fessler1994space}, or compressive sensing to extract the multi-path parameters based on the channel. As such, the process of inferring ${{\mathbf{h}}_{2}}$ from ${{\mathbf{h}}_{1}}$ can be divided into three phases. F{}irstly, according to Assumption 1, obtain the multi-path parameters ${{\alpha }_{l}}\left( {\lambda}_{1}  \right)$, $\theta _{l,1}$, and $\phi _{l,1}$ of the known channel vector ${{\mathbf{h}}_{1}}$. Secondly, infer the multi-path channel parameters ${{\alpha }_{l}}\left( {\lambda}_{2}  \right)$, $\theta _{l,2}$, and $\phi _{l,2}$ of ${{\mathbf{h}}_{2}}$ according to the geometric relationship between ${\rm Panel}_{1}$ and ${\rm Panel}_{2}$. F{}inally, reconstruct the channel vector ${{\mathbf{h}}_{2}}$ based on the inferred multi-path parameters. In fact, Assumption 1 is not a must if instead of estimating the composite channel ${{\mathbf{h}}_{1}}$, the aforementioned multi-path parameters are directly estimated at the f{}irst place for ${\rm Panel}_{1}$. This is usually the case for integrated sensing and communication (ISAC) systems \cite{zhang2023integrated}. Since the other two steps are either well studied or straightforward, this letter mainly focuses on the second step, i.e.,  inferring ${{\alpha }_{l}}\left( {\lambda}_{2}  \right)$, $\theta _{l,2}$, and $\phi _{l,2}$ from ${{\alpha }_{l}}\left( {\lambda}_{1}  \right)$, $\theta _{l,1}$, and $\phi _{l,1}$.
\vspace{-2mm}
\section{Channel Inference Across Panels}
\vspace{-1mm}
To distinguish between far-f{}ield from near-f{}ield scenarios, the classical Rayleigh distance is def{}ined as ${{r}_{Rayl}}=\frac{2{{D}^{2}}}{\lambda }$\cite{selvan2017fraunhofer}, where $D$ is the overall aperture of ${\rm Panel}_{1}$ and ${\rm Panel}_{2}$, and \mbox{$\lambda =\min\left\{ {{\lambda }_{1}},{{\lambda }_{2}} \right\}$}. Denote the distance between the UE and the reference antenna elements of ${\rm Panel}_{1}$ and ${\rm Panel}_{2}$ as ${{R}_{1}}$ and ${{R}_{2}}$, respectively, and let $r=\min\left\{ {{R }_{1}},{{R}_{2}} \right\}$. When $r\ge {{r}_{Rayl}}$, UE is in the far-f{}ield of the antenna panels, and ${\rm Panel}_{1}$ and ${\rm Panel}_{2}$ would share the same multi-path parameters, such as number of multi-paths and their AoDs \cite{HLuandYZeng}; otherwise, UE is in the near-f{}ield and ${\rm Panel}_{1}$ and ${\rm Panel}_{2}$ would have different multi-path parameters\cite{selvan2017fraunhofer}.
\vspace{-4mm}
\subsection{Far-f{}ield Free-Space}
In far-f{}ield free-space scenarios, there is only one line of sight (LoS) path, i.e., $L_{k}=1$, so that the channel vectors in \eqref{eq1} are simplf{}ied to
\begin{equation}
{{\mathbf{h}}_{k}}=\sqrt{{{N}_{k}}}{{\alpha }}\left( {{\lambda}_{k}} \right){{\bf{a}}_{k}}\left( \theta_{k},\phi_{k} \right), k=1, 2.\label{eq5}
\end{equation}

Therefore, the channel vector ${{\mathbf{h}}_{2}}$ can be inferred from ${{\mathbf{h}}_{1}}$ if the parameters ${{\alpha }}\left( {{\lambda}_{2}} \right)$, $\theta_{2}$ and $\phi_{2}$ of ${\rm Panel}_{2}$ can be inferred from ${{\alpha }}\left( {{\lambda}_{1}} \right)$, $\theta_{1}$ and $\phi_{1}$.

As UE is in the far-f{}ield of the overall aperture $D$ formed by both panels, the AoDs for ${\rm Panel}_{1}$ and ${\rm Panel}_{2}$ are equal, so that $\theta_{2}=\theta_{1}$ and $\phi_{2}=\phi_{1}$. Furthermore, based on Friis transmission equation, the complex-valued channel gain $\alpha ({\lambda}_{k})$ can be written as
\setlength\abovedisplayskip{4pt}
\setlength\belowdisplayskip{4pt}
\begin{equation}
\alpha \left( {\lambda}_{k}  \right)=\frac{{\lambda}_{k} }{4\pi R_{k}}{{e}^{-\frac{j2\pi R_{k}}{\lambda_{k} }}}, k=1, 2,\label{eq6}
\end{equation}
where $R_{k}$ is the distance from the UE to the reference antenna element of ${\rm Panel}_{k}$.

Since ${{R}_{1}}$ and ${{R}_{2}}$ are larger than the classical Rayleigh distance, the approximation ${{R}_{1}}\approx {{R}_{2}}$ can be used when evaluating the amplitude of ${{\alpha }}\left( {{\lambda}_{2}} \right)$. However, the phase calculation needs to consider the wavelength and the distance of the signal to ${\rm Panel}_{2}$. It can be shown that ${{\alpha }}\left( {{\lambda}_{2}} \right)$ can be inferred from ${\alpha}({\lambda}_1)$ as
\setlength\abovedisplayskip{-1pt}
\setlength\belowdisplayskip{1pt}
\begin{equation}
\alpha\left(\lambda_{2}\right)=\left|\alpha\left(\lambda_{1}\right)\right| \frac{\lambda_{2}}{\lambda_{1}}\left(\frac{\alpha\left(\lambda_{1}\right)}{\left|\alpha\left(\lambda_{1}\right)\right|}\right)^{\frac{\lambda_{1}}{\lambda_{2}}} e^{\frac{j 2 \pi \Delta d \sin \left(\theta_{2}\right)}{\lambda_{2}}},\label{eq7}
\end{equation}
where $\left| \cdot  \right|$ is the modulus operation for complex numbers, $\Delta d$ is the distance between the reference antenna elements of ${\rm Panel}_{1}$ and ${\rm Panel}_{2}$ as shown in Fig.~\ref{fig1}.

\begin{IEEEproof}
Please refer to Appendix A.
\end{IEEEproof}
Therefore, from the above analysis, we can obtain the following theorem:
\begin{theorem}
In the far-f{}ield free-space scenario, ${{\mathbf{h}}_{2}}$ can be fully inferred from ${{\mathbf{h}}_{1}}$ by using $\theta_{2}=\theta_{1}$, $\phi_{2}=\phi_{1}$, and ${{\alpha }}\left( {{\lambda}_{2}} \right)$ in \eqref{eq7}.\label{Theorem 1}
\end{theorem}
\vspace{-4mm}
\subsection{Near-f{}ield Free-Space}
\begin{figure}[htbp]
\vspace{-8mm}
\centerline{\includegraphics[scale=0.40]{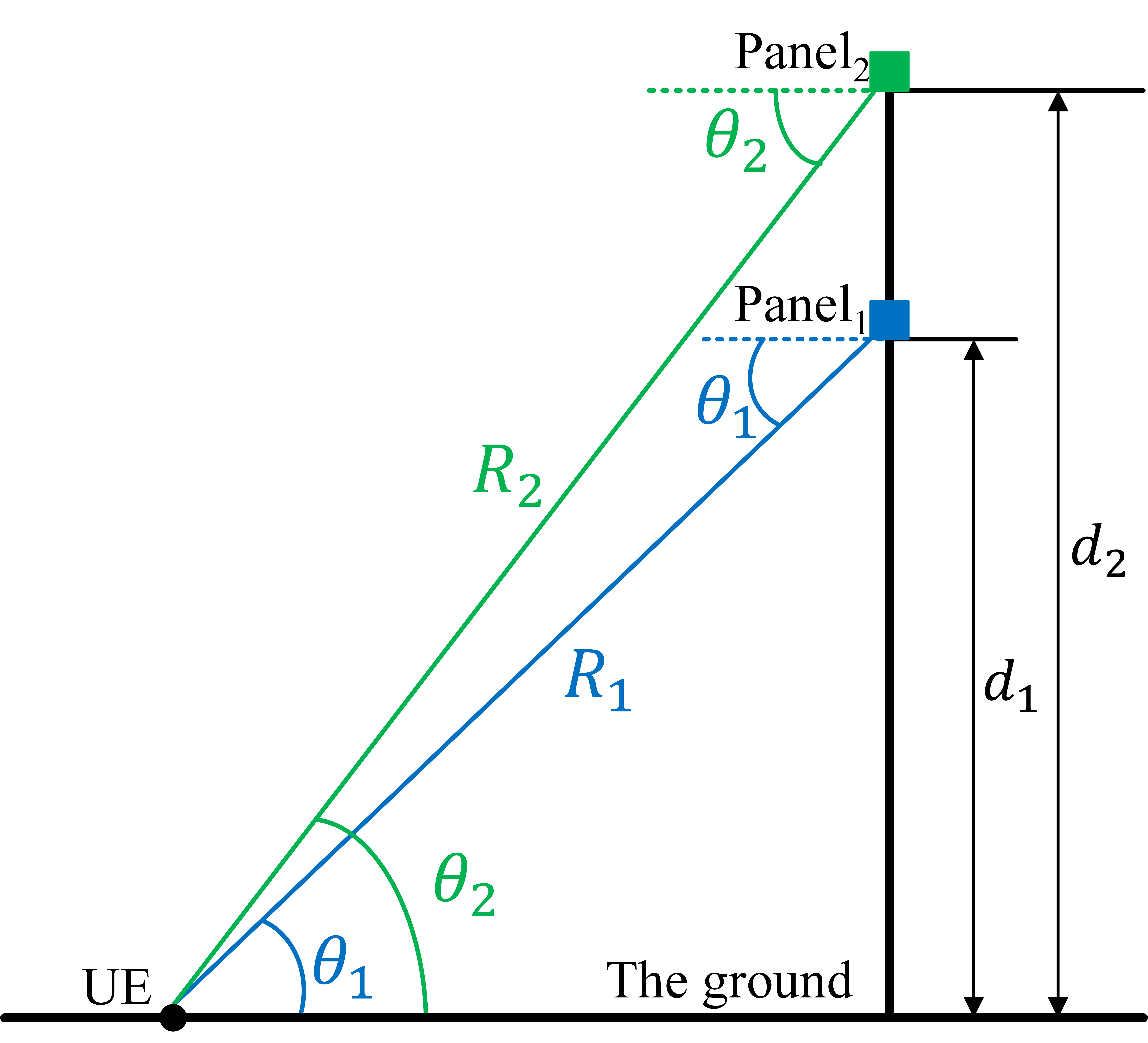}}
\vspace{-2mm}
\caption{Geometric relation between two panels in near-f{}ield free-space .}
\label{fig2}
\end{figure}

When the UE is in the near-f{}ield of the overall aperture $D$, the AoDs of ${\rm Panel}_{1}$ and ${\rm Panel}_{2}$ are different in general, as illustrated in Fig.~\ref{fig2}. By exploiting the geometric relationship between the two panels, it is not diff{}icult to infer the AoD $\theta_{2}$ based on $\theta_{1}$, given by
\setlength\abovedisplayskip{1.5pt}
\setlength\belowdisplayskip{1.5pt}
\begin{equation}
\theta_{2}={{\tan }^{-1}}\left( \frac{{{d}_{2}}}{{{d}_{1}}}{\tan}\theta_{1} \right).\label{eq8}
\end{equation}

Besides, since ${\rm Panel}_{1}$ and ${\rm Panel}_{2}$ are in the same vertical plane, the azimuth AoDs of both panels are equal, i.e., \mbox{$\phi_{2}=\phi_{1}$}.

\vspace{-2mm}
Furthermore, with $\theta_2$ obtained in \eqref{eq8}, the distance ${{{R}}_{2}}$ from UE to reference antenna element of ${{\rm Panel}_{2}}$ can be obtained as
\setlength\abovedisplayskip{-1pt}
\setlength\belowdisplayskip{1pt}
\begin{equation}
{{R}_{2}}=\frac{{{d}_{2}}}{\left|{\sin}\theta_{2}\right|}.\label{eq9}
\end{equation}

Thus, the complex-valued gain ${{\alpha }}\left( {{\lambda}_{2}} \right)$ can be obtained by substituting ${R}_{2}$ into \eqref{eq6}.
\begin{theorem}
In the near-f{}ield free-space scenario, ${{\mathbf{h}}_{2}}$ can be fully inferred from ${{\mathbf{h}}_{1}}$ by using $\phi_{2}=\phi_{1}$, $\theta_{2}$ in \eqref{eq8}, and ${{\alpha }}\left( {{\lambda}_{2}} \right)$ in \eqref{eq6}. \label{Theorem 2}
\end{theorem} 
\vspace{-4mm}
\subsection{Multi-path Sharing Far-f{}ield Scatterers}
\vspace{-1mm}
For multi-path scenarios, let ${{S}_{l}}$ denote the locations of scatterer $l$, which are assumed to be shared by both panels. In the far-f{}ield scenario, ${{S}_{l}}$ are in the far-f{}ield of the overall aperture $D$, so that the AoDs of ${{\rm Panel}_{1}}$ and ${{\rm Panel}_{2}}$ of the multi-paths are equal, i.e., $L_{1}=L_{2}$, and $\theta _{l,2}=\theta _{l,1}$ and $\phi _{l,2}=\phi _{l,1}$, $\forall l$. 

However, it is diff{}icult to infer the path gain $\alpha _{l}\left( {{\lambda}_{2}} \right)$ from $\alpha _{l}\left( {{\lambda}_{1}} \right)$ since the scattered power loss and phase shift when the signal interacts with the scatterers depend on the scatterer surface material in a sophisticated manner. Nevertheless, the inferred angle information is highly valuable to facilitate the estimation of ${{\mathbf{h}}_{2}}$. One possible approach is by combining the inferred angle information with a small amount of real-time training \cite{wu2023environment}.

\begin{theorem}
In the multi-path sharing far-f{}ield scatterers scenario, AoDs of the MPCs of ${{\mathbf{h}}_{2}}$ can be inferred by using $\theta _{l,2}=\theta _{l,1}$, and $\phi _{l,2}=\phi _{l,1}$.\label{Theorem 3}
\end{theorem}
\vspace{-4mm}
\subsection{Multi-path Sharing Near-f{}ield Scatterers}
\vspace{-6mm}
\begin{figure}[htbp]
\centerline{\includegraphics[scale=0.4]{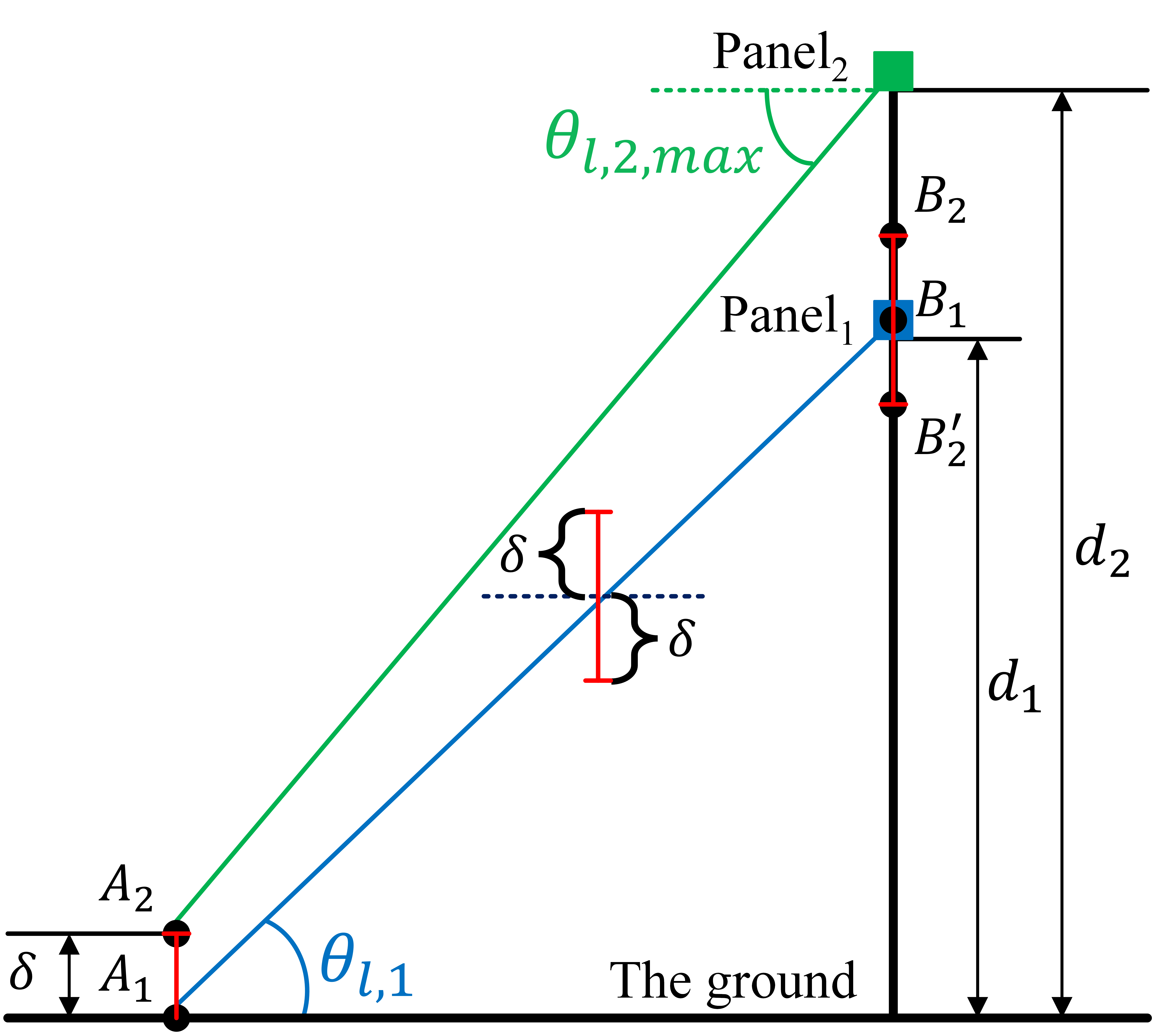}}
\vspace{-2mm}
\caption{Geometric relation between two panels in multi-path scenario sharing near-f{}ield scatterers.}
\label{fig3}
\end{figure}
\vspace{-2mm}
For multi-path scenarios when the scatterers $S_{l}$  are in the near-f{}ield of the overall aperture $D$, the elevation AoDs of the MPCs are in general different for the two panels, as illustrated in  Fig.~\ref{fig3}. Although we cannot determine the exact location of scatterer $S_{l}$, based on the AoD $\theta_{l,1}$ of ${{\rm Panel}_{1}}$, the scattering point of ${\rm Panel}_{1}$ must be located at a point along the line segment $A_{1}B_{1}$ shown in Fig.~\ref{fig3}. Furthermore, in practice, there is usually a slight deviation in the exact scattering point of ${\rm Panel}_{2}$. Specif{}ically, the scattering point of ${\rm Panel}_{2}$ may vertically deviate from that of ${\rm Panel}_{1}$ by a maximum value of $\delta$, as illustrated in Fig.~\ref{fig3}. Therefore, the range of $\theta _{l,2}$ can be determined by considering two extreme cases. Specif{}ically, if the scattering point of ${\rm Panel}_{1}$ is located at the furthest point along direction $\theta_{l,1}$, i.e., point $A_{1}$, and the scattering point of ${\rm Panel}_{2}$ is located at point $A_{2}$. Then the maximum value of $\theta _{l,2}$ is
\begin{equation}
\theta _{l,2,\max }={{\tan }^{-1}}\left( \frac{{{d}_{2}}-\delta}{{{d}_{1}}}{\tan}\theta _{l,1} \right).\label{eq10}
\end{equation}

If the scattering point of ${\rm Panel}_{1}$ is located at point $B_{1}$, it is obvious that $\theta _{l,2,\min }=-\frac{\pi }{2}$, irrespective of the location of the scattering point of ${\rm Panel}_{2}$ along the line segment $B_{2}B_{2}^{'}$ shown in Fig.~\ref{fig3}. Therefore, the range of $\theta _{l,2}$ is $\left ( -\frac{\pi }{2},\left. {{\tan }^{-1}}\left( \frac{{{d}_{2}}-\delta}{{{d}_{1}}}{\tan}\theta _{l,1} \right)\right ]\right.$. It is obvious that the larger $\delta$ is, the wider the range of $\theta _{l,2}$, i.e., less information is inferred for ${\rm Panel}_{2}$. Therefore, in practice, there is a tradeoff for selecting the values of $\delta$.
\begin{theorem}
In the multi-path sharing near-f{}ield scatterers scenario, the azimuth AoDs and the range of elevation AoDs of MPCs of ${{\mathbf{h}}_{2}}$ can be inferred by using $\phi _{l,2}=\phi _{l,1}$, and $\theta _{l,2}\in\left ( -\frac{\pi }{2},\left. {{\tan }^{-1}}\left( \frac{{{d}_{2}}-\delta}{{{d}_{1}}}{\tan}\theta _{l,1} \right)\right ]\right.$.\label{Theorem 4}
\end{theorem}

Based on the above analysis, the results of channel inference across antenna panels under different scenarios are summarized in Table~\ref{table1}. Note that for the most complex scenarios where the multi-path scatterers of different panels are different, it is very challenging to directly infer useful channel knowledge across panels. In this case, the technique of CKM \cite{zeng2021toward}, \cite{zeng2023tutorial} might be useful since it utilizes the location and environment information of the actual propagation environment. 
\vspace{-6mm}
\begin{table}[h]
\caption{Summary of inter-panel channel inference under different scenarios.}
\vspace{-4mm}
\label{table1}
\begin{center}
\renewcommand{\arraystretch}{1.2}
\begin{tabular}{|l|l|}
\hline
\multicolumn{1}{|c|}{\textbf{Scenarios}}                                           & \multicolumn{1}{c|}{\textbf{Inference results}}                                                                           \\ \hline
Far-f{}ield free-space                                                               & \multirow{2}{*}{${\mathbf{h}}_{2}$ can be completely inferred.}                                                                            \\ \cline{1-1}
Near-f{}ield free-space                                                              &                                                                                                                           \\ \hline
\begin{tabular}[c]{@{}l@{}}Multi-path sharing\\ far-f{}ield scatterers\end{tabular}  & $\phi _{l,2}$ and $\theta _{l,2}$ can be inferred.                                                                                                   \\ \hline
\begin{tabular}[c]{@{}l@{}}Multi-path sharing\\ near-f{}ield scatterers\end{tabular} & $\phi_{l,2}$ and the range of $\theta _{l,2}$ can be inferred.                                                                                      \\ \hline
\begin{tabular}[c]{@{}l@{}}Multi-path with\\ different scatterers\end{tabular}     & \begin{tabular}[c]{@{}l@{}}Challenging to directly infer useful channel\\ knowledge across panels. May use CKM.\end{tabular} \\ \hline
\end{tabular}
\end{center}
\end{table}
\vspace{-8mm}
\section{Simulation results}
\vspace{-1mm}
In this section, the discussed channel inference techniques under different scenarios are verif{}ied through ray-tracing simulation. The commercial ray tracing software Remcom Wireless Insite\footnote{\url{https://www.remcom.com/wireless-insite-em-propagation-software}} is used to generate the ground-truth channel information. ${{\rm Panel}_{1}}$ and ${{\rm Panel}_{2}}$ are $16\times 16$ UPAs, with \mbox{${{f}_{1}}=28$ GHz} and \mbox{${{f}_{2}}=39$ GHz}, respectively. The height of ${{\rm Panel}_{1}}$ is f{}ixed to ${{d}_{1}}=15$ meters, while ${{d}_{2}}$ varies among 16, 18 and 20 meters.

\vspace{-4mm}
\subsection{Free-Space scenarios}
\vspace{-1mm}
For both far-f{}ield and near-f{}ield free-space scenarios, ${{\mathbf{h}}_{2}}$ can be in principle fully inferred from ${{\mathbf{h}}_{1}}$. Let ${{\mathbf{h_{2}}}}$ and $\widehat{\mathbf{h}}_{2}$ denote the true and inferred channel vectors, respectively. Then we use the following correlation coeff{}icient to evaluate the quality of inference:
\setlength\abovedisplayskip{1pt}
\setlength\belowdisplayskip{1pt}
\begin{equation}
F=\frac{\lvert\widehat{\mathbf{h}}_{2}^{H} \mathbf{h}_{2}\rvert^{2}}{\lvert\lvert\widehat{\mathbf{h}}_{2}\rvert\rvert^{2}\lvert\lvert\mathbf{h}_{2}\rvert\rvert^{2}}.\label{eq11}
\end{equation}

\begin{figure}[htbp]
\centerline{\includegraphics[scale=0.53]{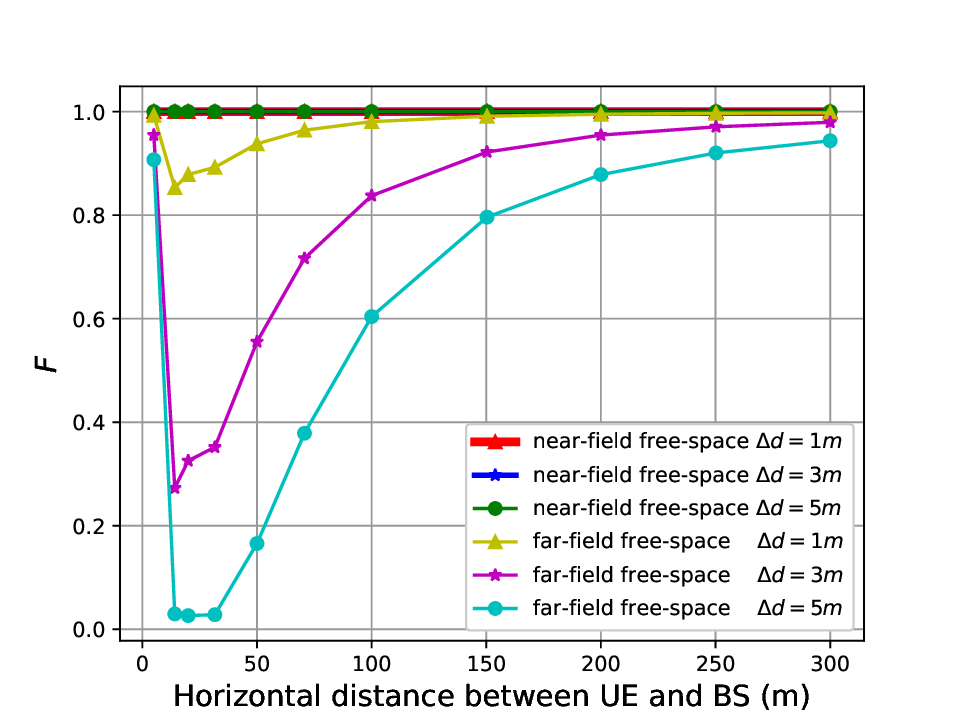}}
\vspace{-2mm}
\caption{Comparison of channel correlation coeff{}icient for the inference methods in far-/near-f{}ield free-space scenario with three different panel spacings. }
\label{fig4} 
\end{figure}

Note that $F$ is a highly relevant metric for channel inference if the inferred result is mainly used for multi-antenna beamforming. Specif{}ically, with Cauchy-Swhartz inequality, it is not diff{}icult to see that $F$ takes the value between 0 and 1, and the higher its value, the better the channel inference for beamforming gain. Fig.~\ref{fig4} shows the correlation coeff{}icient in far-/near-f{}ield free-space scenario with three different panel spacings: 1 m, 3 m, and 5 m. According to ${{r}_{Rayl}}=\frac{2{{D}^{2}}}{\lambda }$, the near-f{}ield range increases with the panel spacing. Therefore, even with an increase in panel spacing, the inference method in near-f{}ield free-space scenario can maintain a good inference quality. However, for the method in far-f{}ield free-space scenario, the overall inference quality drops with an increase in panel spacing. Note that the elevation angle $\theta $ and azimuth angle $\phi $ are the main factors that determine $F$. However, due to the geometric relationship between ${{\rm Panel}_{1}}$ and ${{\rm Panel}_{2}}$, only the value of $\theta $ changes signif{}icantly. Therefore, the change of $F$ mainly depends on the inference accuracy of $\theta $. According to the inference method in the far-f{}ield free-space scenario, we can conclude that the estimated value $\widehat{\theta}_{2}$ of $\theta _{2}$ satisf{}ies $\widehat{\theta}_{2}=\theta _{1}$. If the horizontal distance between UE and BS is $d_{x}$, according to the geometric relationship in Fig.~\ref{fig2}, the inference error of $\theta_{2}$ is
\setlength\abovedisplayskip{1pt}
\setlength\belowdisplayskip{1pt}
\begin{equation}
\begin{split}
\Delta \theta _{2}&=\left|{\theta _{2}}-{\widehat{\theta}_{2}}\right|=\left|{\theta _{2}}-{\theta _{1}}\right|={{\tan }^{-1}}\left(\frac{d_{2}}{{d_{x}}}\right)-{{\tan }^{-1}}\left(\frac{d_{1}}{{d_{x}}}\right)\\ 
&={{\tan }^{-1}}\left(\frac{d_{x}\left(d_{2}-d_{1}\right)}{{d_{x}}^{2}+d_{1}d_{2}}\right),\label{eq12}
\end{split}
\end{equation}
where $\Delta\theta _{2}$ f{}irstly increases and then decreases as $d_{x}$ increases. Therefore, for the inference method in the far-f{}ield free-space scenario, the $F$ decreases f{}irst and then increases.   
\vspace{-5mm}
\vspace{-6mm}
\subsection{Multi-path scenarios}
\begin{figure}[htbp]
\centerline{\includegraphics[scale=0.25]{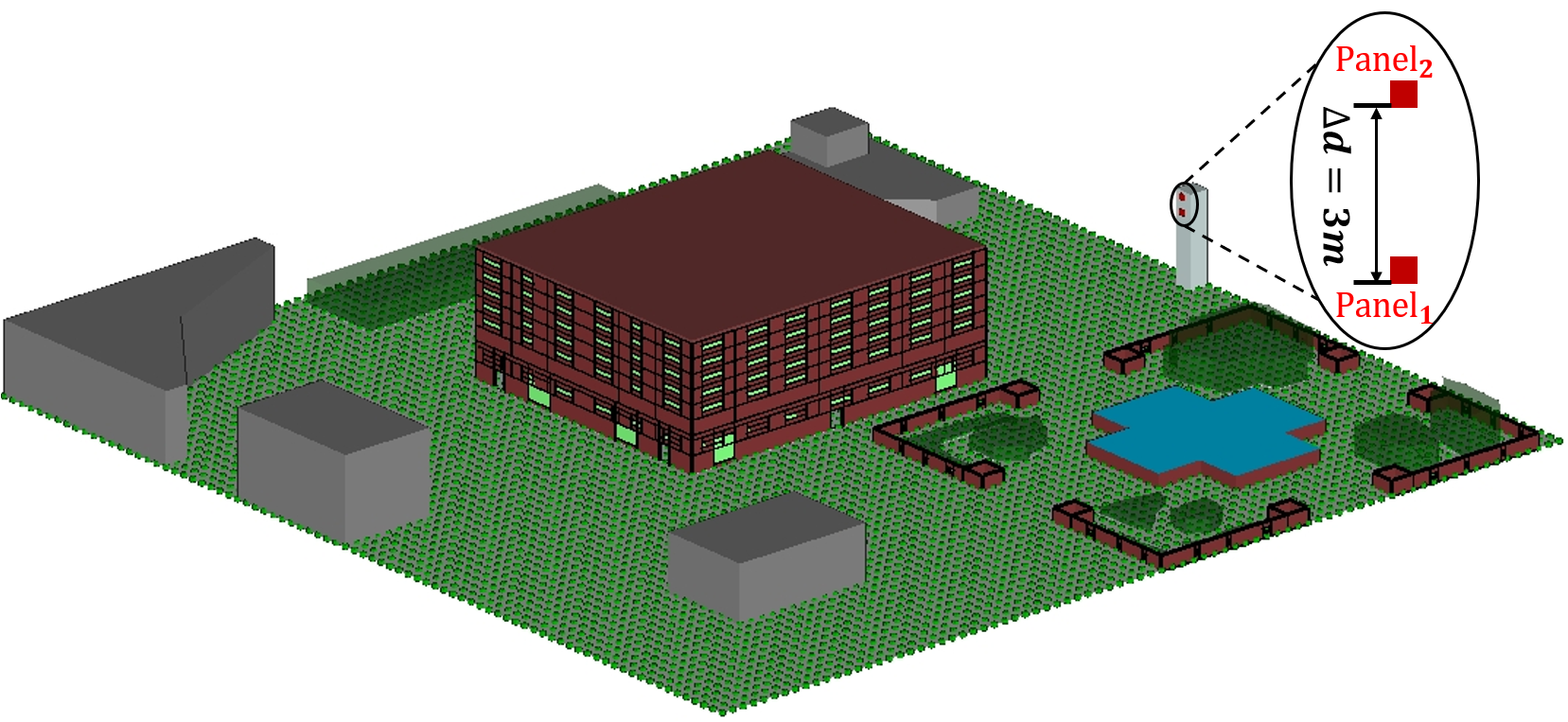}}
\vspace{-4mm}
\caption{Physical environment for two-panel communications. The red little boxes are ${\rm Panel}_{1}$ and ${\rm Panel}_{2}$. The green little squares that cover the entire area are UE locations, and they are spaced 2.5 meters apart. The large red and gray objects are buildings. The green irregular objects are vegetation, and the blue area is the pond.}
\label{fig5} 
\end{figure}

\begin{figure}[htbp]
\centerline{\includegraphics[scale=0.50]{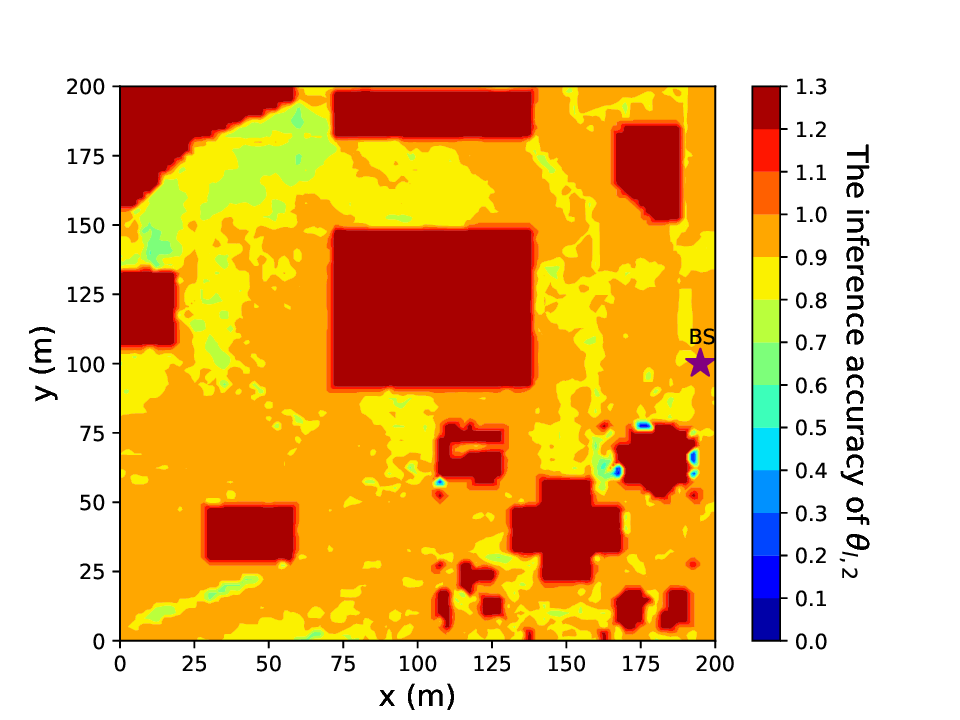}}
\vspace{-3mm}
\caption{The accuracy for inferring $\theta_{l,2}$ for multi-path sharing near-f{}ield scatterers scenario with $\delta=0.15$ meters.}
\label{fig6} 
\end{figure}

Fig.~\ref{fig5} shows the actual diffuse scattering environment of the multi-path scenario, with the diffuse scattering factor set to 0.2. The entire area contains 4848 UE locations, and the maximum number of MPCs is 25 when conf{}iguring Wireless Insite of Remcom. These UEs have a total of 95850 and 93591 MPCs between ${{\rm Panel}_{1}}$ and ${{\rm Panel}_{2}}$, respectively. Consider the MPCs that have common scatterers between ${{\rm Panel}_{1}}$ and ${{\rm Panel}_{2}}$. Then, a total of 85673 MPCs meet this criterion, which account for $89.4\%$ and $91.5\%$ of the total MPCs for ${{\rm Panel}_{1}}$ and ${{\rm Panel}_{2}}$, respectively. In the simulation, we set \mbox{$\delta=0.15$} meters and only consider the 85673 selected MPCs. In the scene depicted in Fig.~\ref{fig5}, utilizing the inference method in multi-path sharing far-f{}ield scatterers scenario to infer the value of $\theta _{l,2}$, i.e. $\theta _{l,2}=\theta _{l,1}$, the average inference error for the entire region is $1.52^{\circ}$. For multi-path sharing near-f{}ield scatterers scenario, since there is no prior information of the exact location of scatterers $S_{l}$, only the range of $\theta_{l,2}$ can be obtained. In Fig.~\ref{fig6}, the purple pentagon represents the position of the BS and the dark red areas are the buildings/vegetation/pond. The value at each UE location is the proportion of the number of MPCs with correctly inferred $\theta _{l,2}$ to the total number of MPCs with common scatterers between ${{\rm Panel}_{1}}$ and ${{\rm Panel}_{2}}$ at that location. It can be observed that utilizing the inference method in the multi-path sharing near-f{}ield scatterers scenario, an accuracy over $90\%$ is achieved for inferring the range of values for $\theta _{l,2}$ across the majority area of the entire region. A few areas with inaccurate inference may be attributed to the complexity of the scattering environment.
\vspace{-4mm}
\vspace{-8mm}

\vspace{-6mm}
\section{Conclusion}
 In this letter, we study the problem of channel inference where the BS is equipped with two antenna panels that may use different frequency bands for communication. Four different scenarios are considered, namely far-f{}ield free-space, near-f{}ield free-space, multi-path sharing far-f{}ield scatterers, and multi-path sharing near-f{}ield scatterers. Simulation results based on ray-tracing in a realistic radio propagation environment demonstrate that the channel of one panel can be in principle fully inferred from that of the other panel for both far-f{}ield and near-f{}ield free-space scenarios. On the other hand, for the multi-path scenarios sharing far-f{}ield or near-f{}ield scatterers, the angles or range of angles of the MPCs can be inferred, respectively.
\vspace{-4mm}
\begin{appendices}
\section{Proof of the inference of ${{\alpha }}\left( {{\lambda}_{2}} \right)$ in \eqref{eq7}}

Based on the expression \eqref{eq6}, $\alpha \left( {\lambda}_{2}  \right)=\frac{{\lambda}_{2} }{4\pi R_{2}}{{e}^{-\frac{j2\pi R_{2}}{\lambda_{2} }}}$, where
\begin{equation}
\frac{\lambda_{2}}{4 \pi R_{2}}=\frac{\lambda_{1}}{4 \pi R_{1}} \frac{\lambda_{2} R_{1}}{\lambda_{1} R_{2}}\approx\frac{\lambda_{1}}{4 \pi R_{1}} \frac{\lambda_{2}}{\lambda_{1}}=\left|\alpha\left(\lambda_{1}\right)\right| \frac{\lambda_{2}}{\lambda_{1}},\label{eq13}
\end{equation}

\begin{equation}
\begin{split}
e^{\frac{-j2\pi R_2}{\lambda_2}}&=e^{\frac{-j2\pi (R_1-\Delta d \sin ({\theta}_{2}))}{\lambda_2}}=(e^{\frac{-j2\pi R_1}{\lambda_1}})^{\frac{\lambda_1}{\lambda_2}}e^{\frac{j2\pi \Delta d \sin ({\theta}_{2})  }{\lambda_2}}\\
&={\left( \frac{\alpha(\lambda_1)}{|\alpha(\lambda_1)|} \right )}^{\frac{{{\lambda}_{1}}}{{{\lambda}_{2}}}}e^{\frac{j2\pi \Delta d \sin ({\theta}_{2})  }{\lambda_2}}.\label{eq14}
\end{split}
\end{equation}

The approximation in equations \eqref{eq13} follows by letting \mbox{${{R}_{1}}\approx {{R}_{2}}$} when evaluating the amplitude for far-f{}ield scenario, and the range of $\theta_{2}$ in equations \eqref{eq14} is $\left(-\frac{\pi}{2},0\right)$. Therefore, the inference of ${{\alpha }}\left( {{\lambda}_{2}} \right)$ in \eqref{eq7} is proved.

\end{appendices}
\vspace{2mm}

\bibliographystyle{IEEEtran}
\bibliography{reference.bib}

\end{document}